\documentclass[nofootinbib,prd]{revtex4}%
\usepackage{amsmath}
\usepackage{amsfonts}
\usepackage{amssymb}
\usepackage{graphicx}%
\begin{document}
\title{The Cosmological Constant as an Eigenvalue of the Hamiltonian constraint in a
Varying Speed of Light theory}
\author{Remo Garattini}
\email{Remo.Garattini@unibg.it}
\affiliation{Universit\`{a} degli Studi di Bergamo, Facolt\`{a} di Ingegneria}
\affiliation{Viale Marconi 5, 24044 Dalmine (Bergamo) Italy}
\affiliation{I.N.F.N. - sezione di Milano, Milan, Italy}
\author{Mariafelicia De Laurentis}
\affiliation{Tomsk State Pedagogical University, ul. Kievskaya, 60, 634061 Tomsk, Russia}
\affiliation{National Research Tomsk State University, Lenin Avenue, 36, 634050 Tomsk,
Russia }

\begin{abstract}
In the framework of a Varying Speed of Light theory, we study the eigenvalues
associated with the Wheeler-DeWitt equation representing the vacuum
expectation values associated with the cosmological constant. We find that the
Wheeler-DeWitt equation for the Friedmann-Lema\^{\i}tre-Robertson-Walker
metric is completely equivalent to a Sturm-Liouville problem provided that the
related eigenvalue and the cosmological constant be identified. The explicit
calculation is performed with the help of a variational procedure with trial
wave functionals related to the Bessel function of the second kind $K_{\nu
}\left(  x\right)  $. We find the existence of a family of eigenvalues
associated to a negative power of the scale. Furthermore, we show that at the
inflationary scale such a family of eigenvalues does not appear.

\end{abstract}
\maketitle


\section{Introduction}

What is the Cosmological Constant? How can be computed? These are some of the
many puzzling questions which are still unsolved. Basically the Cosmological
Constant can be connected to the energy of the vacuum. However, the absence of
a complete Quantum Gravitational theory increases the number of questions
instead of giving answers. General Relativity (GR) is the best theory
explaining the behavior of the gravitational field including also the
cosmological constant. However GR fails to describe the gravitational field in
the quantum range. Despite of this problem, in GR there exists a quantization
procedure known as the Wheeler-De Witt equation (WDW)\cite{DeWitt} which
encodes some aspects of the quantum properties of the gravitational field
included the cosmological constant. We say \textquotedblleft
some\textquotedblright, because a complete solution of the WDW equation does
not exist and one needs to reduce the degree of the difficulty by fixing a
background and freezing some degrees of freedom. The WDW equation is the
quantum version of the classical Hamiltonian constraint representing the
invariance under time reparametrization. Its derivation is a consequence of
the Arnowitt-Deser-Misner (ADM) decomposition \cite{ADM} of space time based
on the following line element%
\begin{equation}
ds^{2}=g_{\mu\nu}\left(  x\right)  dx^{\mu}dx^{\nu}=\left(  -N^{2}+N_{i}%
N^{i}\right)  dt^{2}+2N_{j}dtdx^{j}+g_{ij}dx^{i}dx^{j},
\end{equation}
where $N$ is the lapse function and $N_{i}$ the shift function. In terms of
the ADM variables, the four dimensional scalar curvature $R$ can be decomposed
in the following way%
\begin{equation}
\mathcal{R}=R+K_{ij}K^{ij}-\left(  K\right)  ^{2}-2\nabla_{\mu}\left(
Ku^{\mu}+a^{\mu}\right)  ,\label{R}%
\end{equation}
where%
\begin{equation}
K_{ij}=-\frac{1}{2N}\left[  \partial_{t}g_{ij}-N_{i|j}-N_{j|i}\right]
\end{equation}
is the second fundamental form, $K=$ $g^{ij}K_{ij}$ is its trace, $R$ is the
three dimensional scalar curvature and $\sqrt{g}$ is the three dimensional
determinant of the metric. The last term in $\left(  \ref{R}\right)  $
represents the boundary terms contribution where the four-velocity $u^{\mu}$
is the timelike unit vector normal to the spacelike hypersurfaces
($t$=constant) denoted by $\Sigma_{t}$ and $a^{\mu}=u^{\alpha}\nabla_{\alpha
}u^{\mu}$ is the acceleration of the timelike normal $u^{\mu}$. Thus%
\begin{equation}
\mathcal{L}\left[  N,N_{i},g_{ij}\right]  =\sqrt{-\text{\/\thinspace
\thinspace}^{4}\text{\/{}\negthinspace}g}\left(  \mathcal{R}-2\Lambda\right)
=\frac{N}{2\kappa}\sqrt{g}\text{ }\left[  K_{ij}K^{ij}-K^{2}+\,R-2\Lambda
-2\nabla_{\mu}\left(  Ku^{\mu}+a^{\mu}\right)  \right]  \label{Lag}%
\end{equation}
represents the gravitational Lagrangian density where $\kappa=8\pi G$ with $G$
the Newton's constant and for the sake of generality we have also included a
cosmological constant $\Lambda$. After a Legendre transformation, the WDW
equation simply becomes%
\begin{equation}
\mathcal{H}\Psi=\left[  \left(  2\kappa\right)  G_{ijkl}\pi^{ij}\pi^{kl}%
-\frac{\sqrt{g}}{2\kappa}\!{}\!\left(  \,\!R-2\Lambda\right)  \right]
\Psi=0,\label{WDWO}%
\end{equation}
where $G_{ijkl}$ is the super-metric and where the conjugate super-momentum
$\pi^{ij}$ is defined as%
\begin{equation}
\pi^{ij}=\frac{\delta\mathcal{L}}{\delta\left(  \partial_{t}g_{ij}\right)
}=\left(  g^{ij}K-K^{ij}\text{ }\right)  \frac{\sqrt{g}}{2\kappa}.\label{mom}%
\end{equation}
Note that $H=0$, represents the classical constraint which guarantees the
invariance under time reparametrization. The other classical constraint
represents the invariance by spatial diffeomorphism and it is described by
$\pi_{|j}^{ij}=0$, where the vertical stroke \textquotedblleft%
$\vert$%
\textquotedblright\ denotes the covariant derivative with respect to the $3D$
metric $g_{ij}$. Solving Eq.$\left(  \ref{WDWO}\right)  $ allows to extract
information on the early universe and on the cosmological constant. Of course,
the form of the solution is depending on the background one considers. In this
paper, we fix our attention on the Friedmann-Lema\^{\i}tre-Robertson-Walker
(FLRW) metric without matter fields. To the reader, this choice could seem a
restriction, however one has to think that in the very early universe, before
the inflationary phase, it is likely that all the quantum information can be
carried on by the gravitational field, because of its non-linear nature.
However, even in this simplified vision many problems arise, especially for
the inflationary epoch. In recent years, the idea of modifying GR to cure some
of its diseases has been considered. From one side, the so-called $f\left(
R\right)  $ theories have been taken under examination to cure some problems
in the infrared (IR) region\cite{f(R)} and on the other hand modifications on
short scales allowing a power-counting ultraviolet (UV) renormalizable have
been proposed by Ho\v{r}ava motivated by the Lifshitz theory in solid state
physics\cite{Horava}\cite{Lifshitz}. This theory is dubbed as
Ho\v{r}ava-Lifshitz (HL) theory and should recover general relativity in the
IR limit. Nevertheless the price to pay to obtain a renormalizable theory in
the HL\ proposal is that we have no general covariance or, in other words
Lorentz symmetry is broken. Another proposal which distorts gravity in the UV
is Gravity's Rainbow\cite{MagSmo}. Gravity's Rainbow has some appealing
features to explain inflation\cite{AM}. In a series of papers, one of us used
Gravity's Rainbow to cure some divergences appearing in Zero Point Energy
(ZPE) calculations, at least to one loop\cite{GMLM}. The final ZPE result has
been interpreted as an induced Cosmological Constant obtained as an eigenvalue
of an appropriate Sturm-Liouville problem\footnote{See Ref.\cite{GMLM1} for
other applications in the context of Gravity's Rainbow.}. It is interesting to
note that the same idea has been applied in a HL theory\cite{RemoHL}%
\footnote{See also Ref.\cite{OBCZ,AAEEAY}. See also Ref.\cite{RGES} to see how
Gravity's Rainbos and HL theory can be connected.} in a FLRW background, the
final result is that non trivial eigenvalues have been found depending on the
parameters of the theory. Note that in GR, as we will show extensively in the
next section, the cosmological constant cannot be considered as an eigenvalue
of any Sturm-Liouville problem for the FLRW background in a mini-superspace
approximation without matter fields. It appears therefore, that distortions of
GR allow new results that otherwise should not be possible. It remains to
consider another distortion connected with the previous ones: a Varying Speed
of Light (VSL) theory\cite{harko,moffat,Albrecht,Barrow1,Barrow2,Barrow3}. In
this approach, one allows the speed of light to change in some specified way,
in an attempt to solve the major cosmological issues of modern theoretical
physics. It is well known, that one of the major features of Einstein's theory
of relativity is that the speed of light in a vacuum is always at constant
rate. However, the cosmological problems that led to the theoretical
introduction of dark matter and dark energy into modern cosmology have
motivated some physicists to look for solutions in other directions, included
the variation of the speed of light. In VSL, it is supposed that light travels
faster in the early periods of the existence of the Universe and for this
reason, it could solve problems related to the inflationary phase (flatness,
horizon, homogeneity, etc.\ldots)\cite{Barrow1}%
\cite{Kolb90,guth,linde,AA,veneziano}. Of course, this hypothesis breaks the
Lorentz invariance. The VSL model has been embedded within the general
framework of the time varying fine structure constant theory and reformulated
as a dielectric vacuum theory \cite{Barrow2}. Moreover, isotropy and
homogeneity problems may find their appropriate solutions through this
mechanism \cite{ellis,magueio,moff}. Recently quantum cosmological aspect of
VSL models have been studied to see if the \textquotedblleft Tunneling from
Nothing\textquotedblright\cite{Vilenkin}\ and the \textquotedblleft
Hartle-Hawking No-boundary proposal\textquotedblright\cite{HH}\ can be better
approached in this context\cite{harko1,sho}. The purpose of this paper is to
repeat the calculation of Ref.\cite{RemoHL} in a VSL context to see if there
are non trivial eigenvalues of an appropriate Sturm-Liouville problem, which
will be interpreted as a Cosmological Constant induced by quantum fluctuations
of the scale factor. The paper is organized as follows. In Sec. \ref{p1} we
discuss the Wheeler-deWitt equations for
Friedmann-Lema\^{\i}tre-Robertson-Walker space time. While, in Sec. \ref{p2},
we show how it is possible to derive the Wheeler-deWitt equations for
Friedmann-Lema\^{\i}tre-Robertson-Walker space time in presence of varying
speed of light. Conclusions are drawn in Sec.\ref{p3}.

\section{The Wheeler-DeWitt equation for the
Friedmann-Lema\^{\i}tre-Robertson-Walker space-time}

\label{p1}

A homogeneous, isotropic and closed universe is represented by the FLRW line
element%
\begin{equation}
ds^{2}=-N^{2}dt^{2}+a^{2}\left(  t\right)  d\Omega_{3}^{2}, \label{FRW}%
\end{equation}
where%
\begin{equation}
d\Omega_{3}^{2}=\gamma_{ij}dx^{i}dx^{j} \label{domega}%
\end{equation}
is the line element on the three-sphere, $N$ is the lapse function and $a(t)$
denotes the scale factor. Let us consider a very simple mini-superspace model
described by the metric of Eq.$\left(  \ref{FRW}\right)  $. In this
background, the Ricci curvature tensor and the scalar curvature read simply%
\begin{equation}
R_{ij}=\frac{2}{a^{2}\left(  t\right)  }\gamma_{ij}\qquad\mathrm{and}\qquad
R=\frac{6}{a^{2}\left(  t\right)  }~,
\end{equation}
respectively. The Einstein-Hilbert action in $(3+1)$-dim is%
\begin{equation}
S=\frac{1}{16\pi G}\int_{\Sigma\times I}\mathcal{L}~dt~d^{3}x=\frac{1}{16\pi
G}\int_{\Sigma\times I}N\sqrt{g}\left[  K^{ij}K_{ij}-K^{2}+R-2\Lambda\right]
~dt~d^{3}x~, \label{action}%
\end{equation}
with $\Lambda$ the cosmological constant, $K_{ij}$ the extrinsic curvature and
$K$ its trace. Using the line element, Eq.~$\left(  \ref{FRW}\right)  $, the
above written action, Eq.~$\left(  \ref{action}\right)  $, becomes%
\begin{equation}
S=-\frac{3\pi}{4G}\int_{I}\left[  \dot{a}^{2}a-a+\frac{\Lambda}{3}%
a^{3}\right]  dt~,
\end{equation}
where we have computed the volume associated to the three-sphere, namely
$V_{3}=2\pi^{2}$, and set $N=1$.

The canonical momentum reads%
\begin{equation}
\pi_{a}=\frac{\delta S}{\delta\dot{a}}=-\frac{3\pi}{2G}\dot{a}a~,
\end{equation}
and the resulting Hamiltonian density is%
\begin{align}
\mathcal{H} &  =\pi_{a}\dot{a}-\mathcal{L}\nonumber\\
&  =-\frac{G}{3\pi a}\pi_{a}^{2}-\frac{3\pi}{4G}a+\frac{3\pi}{4G}\frac
{\Lambda}{3}a^{3}~.\label{H0}%
\end{align}
Following the canonical quantization prescription, we promote $\pi_{a}$ to a
momentum operator, setting%
\begin{equation}
\pi_{a}^{2}\rightarrow-a^{-q}\left[  \frac{\partial}{\partial a}a^{q}%
\frac{\partial}{\partial a}\right]  ,\label{ordering}%
\end{equation}
where we have introduced a factor order ambiguity $q$. The generalization to
$k=0,-1$ is straightforward. The WDW equation for such a metric is%
\begin{gather}
H\Psi\left(  a\right)  =\left[  -a^{-q}\left(  \frac{\partial}{\partial
a}a^{q}\frac{\partial}{\partial a}\right)  +\frac{9\pi^{2}}{4G^{2}}\left(
a^{2}-\frac{\Lambda}{3}a^{4}\right)  \right]  \Psi\left(  a\right)
\,,\nonumber\\
\left[  -\frac{\partial^{2}}{\partial a^{2}}-\frac{q}{a}\frac{\partial
}{\partial a}+\frac{9\pi^{2}}{4G^{2}}\left(  a^{2}-\frac{\Lambda}{3}%
a^{4}\right)  \right]  \Psi\left(  a\right)  =0.\label{WDW_0}%
\end{gather}
It represents the quantum version of the invariance with respect to time
reparametrization. If we define the following reference length $a_{0}%
=\sqrt{3/\Lambda}$, then Eq.$\left(  \ref{WDW_0}\right)  $ assumes the
familiar form of a one-dimensional Schr\"{o}dinger equation for a particle
moving in the potential%
\begin{equation}
U\left(  a\right)  =\frac{9\pi^{2}a_{0}^{2}}{4G^{2}}\left[  \left(  \frac
{a}{a_{0}}\right)  ^{2}-\left(  \frac{a}{a_{0}}\right)  ^{4}\right]
\,,\label{U(a)}%
\end{equation}
with zero total energy. The potential $U\left(  a\right)  $ resembles a
potential well which is unbounded from below.\textbf{ }When $0<a<a_{0}$,
Eq.~$\left(  \ref{U(a)}\right)  $ implies $U\left(  a\right)  >0$, which is
the classically forbidden region, while for $a>a_{0}$, one gets $U\left(
a\right)  <0$, which is the classically allowed region. It is interesting to
note that for for the special case of the operator ordering $q=-1$, one can
determine exact solution\cite{Vilenkin}. This can be easily verified by
introducing the function%
\[
z\left(  a\right)  =\left(  \frac{3\pi a_{0}^{2}}{4G}\right)  ^{\frac{2}{3}%
}\left(  1-\frac{a^{2}}{a_{0}^{2}}\right)  =z_{0}\left(  1-\frac{a^{2}}%
{a_{0}^{2}}\right)  ,
\]
where the solution can be written in terms of Airy functions, namely%
\begin{equation}
\Psi\left(  a\right)  =\alpha Ai\left(  z\right)  +\beta Bi\left(  z\right)
.\label{Airy}%
\end{equation}
However, the wave function $\left(  \ref{Airy}\right)  $, cannot be normalized
in the following sense%
\begin{equation}
\int_{0}^{\infty}daa^{q}\Psi^{\ast}\left(  a\right)  \Psi\left(  a\right)
.\label{Norm}%
\end{equation}
The same happens for the other special value $q=3$. Even if the WDW equation
$\left(  \ref{WDW_0}\right)  $ has a zero energy eigenvalue, it also has a
hidden structure. Indeed Eq.$\left(  \ref{WDW_0}\right)  $ has the structure
of a Sturm-Liouville eigenvalue problem with the cosmological constant as
eigenvalue. We recall to the reader that a Sturm-Liouville differential
equation is defined by%
\begin{equation}
\frac{d}{dx}\left(  p\left(  x\right)  \frac{dy\left(  x\right)  }{dx}\right)
+q\left(  x\right)  y\left(  x\right)  +\lambda w\left(  x\right)  y\left(
x\right)  =0\label{SL}%
\end{equation}
and the normalization is defined by%
\begin{equation}
\int_{a}^{b}dxw\left(  x\right)  y^{\ast}\left(  x\right)  y\left(  x\right)
.
\end{equation}
In the case of the FLRW model we have the following correspondence%
\begin{align}
p\left(  x\right)   &  \rightarrow a^{q}\left(  t\right)  \,,\nonumber\\
q\left(  x\right)   &  \rightarrow\left(  \frac{3\pi}{2G}\right)  ^{2}%
a^{q+2}\left(  t\right)  \,,\nonumber\\
w\left(  x\right)   &  \rightarrow a^{q+4}\left(  t\right)  \,,\nonumber\\
y\left(  x\right)   &  \rightarrow\Psi\left(  a\right)  \,,\nonumber\\
\lambda &  \rightarrow\frac{\Lambda}{3}\left(  \frac{3\pi}{2G}\right)  ^{2}\,,
\end{align}
and the normalization becomes%
\begin{equation}
\int_{0}^{\infty}daa^{q+4}\Psi^{\ast}\left(  a\right)  \Psi\left(  a\right)
.\label{Norm1}%
\end{equation}
It is a standard procedure, to convert the Sturm-Liouville problem $\left(
\ref{SL}\right)  $ into a variational problem of the form\footnote{Actually
the standard variational procedure prefers the following form%
\begin{equation}
F\left[  y\left(  x\right)  \right]  =\frac{-\left[  y^{\ast}\left(  x\right)
p\left(  x\right)  \frac{d}{dx}y\left(  x\right)  \right]  _{a}^{b}+\int
_{a}^{b}dxp\left(  x\right)  \left(  \frac{d}{dx}y\left(  x\right)  \right)
^{2}-q\left(  x\right)  y\left(  x\right)  }{\int_{a}^{b}dxw\left(  x\right)
y^{\ast}\left(  x\right)  y\left(  x\right)  }\,,
\end{equation}
with appropriate boundary conditions.}%
\begin{equation}
F\left[  y\left(  x\right)  \right]  =\frac{-\int_{a}^{b}dxy^{\ast}\left(
x\right)  \left[  \frac{d}{dx}\left(  p\left(  x\right)  \frac{d}{dx}\right)
+q\left(  x\right)  \right]  y\left(  x\right)  }{\int_{a}^{b}dxw\left(
x\right)  y^{\ast}\left(  x\right)  y\left(  x\right)  }\,.\label{Funct}%
\end{equation}
with boundary condition to be specified. If $y\left(  x\right)  $ is an
eigenfunction of $\left(  \ref{SL}\right)  $, then%
\begin{equation}
\lambda=\frac{-\int_{a}^{b}dxy^{\ast}\left(  x\right)  \left[  \frac{d}%
{dx}\left(  p\left(  x\right)  \frac{d}{dx}\right)  +q\left(  x\right)
\right]  y\left(  x\right)  }{\int_{a}^{b}dxw\left(  x\right)  y^{\ast}\left(
x\right)  y\left(  x\right)  }\,,
\end{equation}
is the eigenvalue, otherwise%
\begin{equation}
\lambda_{1}=\min_{y\left(  x\right)  }\frac{-\int_{a}^{b}dxy^{\ast}\left(
x\right)  \left[  \frac{d}{dx}\left(  p\left(  x\right)  \frac{d}{dx}\right)
+q\left(  x\right)  \right]  y\left(  x\right)  }{\int_{a}^{b}dxw\left(
x\right)  y^{\ast}\left(  x\right)  y\left(  x\right)  }\,.
\end{equation}
\textbf{ }The minimum of the functional $F\left[  y\left(  x\right)  \right]
$ corresponds to a solution of the Sturm-Liouville problem $\left(
\ref{SL}\right)  $ with the eigenvalue $\lambda.$ In the mini-superspace
approach with a FLRW background, one finds \cite{RemoHL}%
\begin{equation}
\frac{\int\mathcal{D}aa^{q}\Psi^{\ast}\left(  a\right)  \left[  -\frac
{\partial^{2}}{\partial a^{2}}-\frac{q}{a}\frac{\partial}{\partial a}%
+\frac{9\pi^{2}}{4G^{2}}a^{2}\right]  \Psi\left(  a\right)  }{\int
\mathcal{D}aa^{q+4}\Psi^{\ast}\left(  a\right)  \Psi\left(  a\right)  }%
=\frac{3\Lambda\pi^{2}}{4G^{2}},\label{WDW_1}%
\end{equation}
The best form of the trial wave function can be guessed by looking the
asymptotic behavior of Eq.$\left(  \ref{WDW_0}\right)  $. For $a\rightarrow
\infty$, we find%
\begin{equation}
\left[  -\frac{\partial^{2}}{\partial a^{2}}-\frac{q}{a}\frac{\partial
}{\partial a}+\frac{9\pi^{2}}{4G^{2}}\left(  a^{2}-\frac{\Lambda}{3}%
a^{4}\right)  \right]  \Psi\left(  a\right)  \simeq\left[  -\frac{\partial
^{2}}{\partial a^{2}}-\frac{3\Lambda\pi^{2}}{4G^{2}}a^{4}\right]  \Psi\left(
a\right)  =0\label{asy}%
\end{equation}
and when $a\rightarrow0$, we find%
\begin{equation}
\left[  -\frac{\partial^{2}}{\partial a^{2}}-\frac{q}{a}\frac{\partial
}{\partial a}+\frac{9\pi^{2}}{4G^{2}}\left(  a^{2}-\frac{\Lambda}{3}%
a^{4}\right)  \right]  \Psi\left(  a\right)  \simeq\left[  -\frac{\partial
^{2}}{\partial a^{2}}-\frac{q}{a}\frac{\partial}{\partial a}+\frac{9\pi^{2}%
}{4G^{2}}a^{2}\right]  \Psi\left(  a\right)  =0.
\end{equation}
When $a\rightarrow0$, the previous equation can be exactly solved by a
superposition of modified Bessel functions of the first $\left(  I_{\nu
}\left(  x\right)  \right)  $ and second kind $\left(  K_{\nu}\left(
x\right)  \right)  $. We find\cite{Wiltshire}%

\begin{equation}
\Psi_{0}\left(  a\right)  =C_{1}{a}^{\left(  1-q\right)  /2}I_{\left(
q-1\right)  /4}{\left(  \frac{3\pi}{4G}{a}^{2}\right)  }+C_{2}{a}^{\left(
1-q\right)  /2}K_{\left(  q-1\right)  /4}{\left(  \frac{3\pi}{4G}{a}%
^{2}\right)  .}\label{sol}%
\end{equation}
However, the solution $\left(  \ref{sol}\right)  $ is exact for a vanishing
eigenvalue. Since our purpose is the evaluation of Eq.$\left(  \ref{WDW_1}%
\right)  $, we need a solution for $a\rightarrow0$, which considers a generic
not vanishing eigenvalue. This is described by%
\begin{gather}
\Psi\left(  a\right)  =C_{1}\exp\left(  -\frac{3\pi a^{2}}{8G}\right)
M{\left(  \frac{q+1}{4}-\frac{GE^{2}}{3\pi},\frac{q+1}{2},\frac{3\pi}{4G}%
{a}^{2}\right)  }\nonumber\\
+C_{2}\exp\left(  -\frac{3\pi a^{2}}{8G}\right)  U{\left(  \frac{q+1}{4}%
-\frac{GE^{2}}{3\pi},\frac{q+1}{2},\frac{3\pi}{4G}{a}^{2}\right)
,}\label{twf}%
\end{gather}
where $M{\left(  a,b,x\right)  }$ and $U{\left(  a,b,x\right)  }$ are the
Kummer functions. For practical purposes, it is useful to transform $M{\left(
a,b,x\right)  }$ and $U{\left(  a,b,x\right)  }$ in terms of $I_{\nu}\left(
x\right)  $ and $K_{\nu}\left(  x\right)  $. We find%
\begin{equation}
\left\{
\begin{array}
[c]{c}%
M{\left(  a+1/2,2a+1,2x\right)  =\Gamma}\left(  1+a\right)  \mathrm{\exp
}\left(  x\right)  I_{a}{\left(  x\right)  /}\left(  x/2\right)  ^{a}\\
\\
U{\left(  a+1/2,2a+1,2x\right)  =}\mathrm{\exp}\left(  x\right)  K_{a}{\left(
x\right)  /}\left(  \sqrt{\pi}\left(  2x\right)  ^{a}\right)
\end{array}
\right.  .\label{Kummer}%
\end{equation}
Since $M{\left(  a,b,x\right)  }$ is proportional to $I_{a}{\left(  x\right)
}$ which is divergent for large $x$, we will fix $C_{1}=0$ to obtain
normalizable solutions. Thus, we consider the following form%
\begin{equation}
\Psi\left(  a\right)  =\exp\left(  -\frac{\beta{a}^{2}}{2}\right)  U{\left(
\frac{q+1}{4},\frac{q+1}{2},\beta{a}^{2}\right)  =\frac{\left(  \beta{a}%
^{2}\right)  ^{\left(  1-q\right)  /4}}{\sqrt{\pi}}K_{\left(  q-1\right)
/4}{\left(  \frac{\beta{a}^{2}}{2}\right)  }},\label{Psi}%
\end{equation}
for the trial wave function and we plug $\left(  \ref{Psi}\right)  $ into
Eq.$\left(  \ref{WDW_1}\right)  $. After an integration over the scale factor
$a\left(  t\right)  $, one gets%
\begin{equation}
\frac{3\Lambda\left(  \beta\right)  \pi^{2}}{4G^{2}}=\frac{\int_{0}^{\infty
}dxxK_{\left(  q-1\right)  /4}^{2}{\left(  x\right)  }}{2\int_{0}^{\infty
}dxx^{2}K_{\left(  q-1\right)  /4}^{2}{\left(  x\right)  }}\left(  -\beta
^{3}+\frac{9\pi^{2}}{4G^{2}}\beta\right)  ,\label{LambdaB}%
\end{equation}
where $\beta$ is a variational parameter and where we have rescaled the
integrals with the help of the results of Appendix \ref{Appe1}. By imposing
that $\Lambda\left(  \beta\right)  $ be stationary against arbitrary
variations of the parameter $\beta$, we obtain%
\begin{equation}
\frac{d}{d\beta}\Lambda\left(  \beta\right)  =\frac{3\int_{0}^{\infty
}dxxK_{\left(  q-1\right)  /4}^{2}{\left(  x\right)  }}{2\int_{0}^{\infty
}dxx^{2}K_{\left(  q-1\right)  /4}^{2}{\left(  x\right)  }}\left(  \frac
{2G}{3\pi}\right)  ^{2}\frac{d}{d\beta}\left(  -\beta^{3}+\left(  \frac{3\pi
}{2G}\right)  ^{2}\beta\right)  =\,0.
\end{equation}
This implies%
\begin{equation}
\beta_{\pm}=\pm\frac{\sqrt{3}\pi}{2G}.\label{sol12}%
\end{equation}
Plugging $\left(  \ref{sol12}\right)  $ into Eq.$\left(  \ref{LambdaB}\right)
$, one finds%
\begin{equation}
\Lambda\left(  \beta_{\pm}\right)  =4\beta_{\pm}\frac{\Gamma\left(  \frac
{3+q}{4}\right)  \Gamma\left(  \frac{5-q}{4}\right)  }{\Gamma\left(
\frac{5+q}{4}\right)  \Gamma\left(  \frac{7-q}{4}\right)  }.
\end{equation}
It is easy to check that $\beta_{+}$ is a maximum and $\beta_{-}$ is a
minimum. However $\beta_{-}$ is negative independently on $q$ and this leads
to a normalization $\left(  \ref{Norm1}\right)  $ in the range $\left(
-\infty,0\right]  $ which is non physical. A further exploration with a pure
Gaussian choice, namely%
\begin{equation}
\Psi\left(  a\right)  =\exp\left(  -\frac{\beta{a}^{2}}{2}\right)
\label{Gauss}%
\end{equation}
for $q=0$, leads to%
\begin{equation}
3\Lambda\left(  \frac{\pi}{2G}\right)  ^{2}=\frac{\int\mathcal{D}a\Psi^{\ast
}\left(  a\right)  \left[  -\frac{\partial^{2}}{\partial a^{2}}+\left(
\frac{3\pi}{2G}\right)  ^{2}a^{2}\right]  \Psi\left(  a\right)  }%
{\int\mathcal{D}aa^{4}\Psi^{\ast}\left(  a\right)  \Psi\left(  a\right)
}=\frac{2}{3}\left(  \beta^{3}+\left(  \frac{3\pi}{2G}\right)  ^{2}%
\beta\right)  .
\end{equation}
The application of the variational procedure leads to imaginary solutions and
therefore it will be discarded. It remains to test the following assumption%
\begin{equation}
\Psi\left(  a\right)  =\exp\left(  -\frac{\beta{a}^{4}}{2}\right)  ,\label{a4}%
\end{equation}
suggested by the asymptotic behavior $\left(  \ref{asy}\right)  $. In the next
section we will discuss the choice $\left(  \ref{a4}\right)  $ as a particular
case of a VSL theory. One could insist in this direction and try to explore
other trial wave functions. However, choices $\left(  \ref{twf}\right)  $,
$\left(  \ref{Gauss}\right)  $ and $\left(  \ref{a4}\right)  $ have been
chosen following the standard procedure for a variational approach. Therefore,
the other choices can only be small variations of the proposed trial wave
functions above mentioned. Therefore we are led to consider a distorted
version of the gravitational field induced by a VSL theory.

\section{The Wheeler-DeWitt equation for the
Friedmann-Lemaitre-Robertson-Walker space-time in the presence of varying
speed of light}

\label{p2}A VSL cosmology model is described by the following line element%
\begin{equation}
ds^{2}=-N^{2}\left(  t\right)  c^{2}\left(  t\right)  dt^{2}+a^{2}\left(
t\right)  d\Omega_{3}^{2},\label{FRWc}%
\end{equation}
where $d\Omega_{3}^{2}$ is described by Eq.$\left(  \ref{domega}\right)  $ and
where $c\left(  t\right)  $ is an arbitrary function of time with the
dimensions of a $\left[  length/time\right]  $. The form of the background is
such that the shift function $N^{i}$ vanishes. Thus, the extrinsic curvature
reads%
\begin{equation}
K_{ij}=-\frac{\dot{g}_{ij}}{2N\left(  t\right)  c\left(  t\right)  }%
=-\frac{\dot{a}\left(  t\right)  }{N\left(  t\right)  c\left(  t\right)
a\left(  t\right)  }g_{ij},\label{Kij}%
\end{equation}
where the dot denotes differentiation with respect to time $t$. The
gravitational action fulfilling the Einstein's Field equation with the speed
of light explicitly written is%
\begin{equation}
S=\frac{1}{16\pi G}\int_{\mathcal{M}}c^{4}\left(  t\right)  \sqrt{-g}Rd^{4}x,
\end{equation}
where we have used the following relationship\textbf{ }$dx^{0}=c\left(
t\right)  dt$. It is easy to write the form of the reduced action of the
mini-superspace. Indeed, reintroducing the speed of light into the action
$\left(  \ref{action}\right)  $, one gets in $(3+1)$%
\begin{equation}
S=\frac{1}{16\pi G}\int_{\Sigma\times I}N\left(  t\right)  c^{4}\left(
t\right)  \sqrt{g}\left[  K^{ij}K_{ij}-K^{2}+R-2\Lambda\right]  ~dt~d^{3}%
x~,\label{actionc}%
\end{equation}
Using the line element, Eq.~$\left(  \ref{FRWc}\right)  $, the above written
action, Eq.~$\left(  \ref{actionc}\right)  $, becomes%
\begin{equation}
S=-\frac{3\pi}{4G}\int_{I}c^{2}\left(  t\right)  \left[  \dot{a}^{2}%
a-ac^{2}\left(  t\right)  +\frac{\Lambda}{3}a^{3}c^{2}\left(  t\right)
\right]  dt~,
\end{equation}
where we have computed the volume associated to the three-sphere, namely
$V_{3}=2\pi^{2}$, and set $N=1$. The canonical momentum reads%
\begin{equation}
\pi_{a}=\frac{\delta S}{\delta\dot{a}}=-\frac{3\pi}{2G}\dot{a}\,a\,c^{2}%
\left(  t\right)  ~,
\end{equation}
and the resulting Hamiltonian density is%
\begin{align}
\mathcal{H} &  =\pi_{a}\dot{a}-\mathcal{L}\nonumber\\
&  =-\frac{G}{3\pi ac^{2}\left(  t\right)  }\pi_{a}^{2}-\frac{3\pi}%
{4G}a\,c^{4}\left(  t\right)  +\frac{3\pi}{4G}\frac{\Lambda}{3}a^{3}%
c^{4}\left(  t\right)  ~.
\end{align}
According to the usual prescription where $\pi_{a}$ is promoted to an
operator, we can write%
\begin{equation}
\pi_{a}\rightarrow-i\hbar c\left(  t\right)  \frac{\partial}{\partial a}%
\end{equation}
and introducing the factor ordering ambiguity%
\begin{equation}
\pi_{a}^{2}\rightarrow-\left(  \hbar c\left(  t\right)  \right)  ^{2}%
a^{-q}\frac{\partial}{\partial a}a^{q}\frac{\partial}{\partial a},
\end{equation}
the WDW equation $\mathcal{H}\Psi=0$ simply becomes%
\begin{equation}
\left[  -\frac{G}{3\pi ac^{2}\left(  t\right)  }\pi_{a}^{2}-\frac{3\pi}%
{4G}a\,c^{4}\left(  t\right)  +\frac{3\pi}{4G}\frac{\Lambda}{3}a^{3}%
c^{4}\left(  t\right)  \right]  \Psi\left(  a\right)  =0.
\end{equation}
Following\cite{Barrow1,Barrow2,Barrow3}, we assume that%
\begin{equation}
c\left(  t\right)  =c_{0}\left(  \frac{a\left(  t\right)  }{a_{0}}\right)
^{\alpha}\label{c(t)}%
\end{equation}
where $a_{0}$ is a reference length scale whose value will be fixed later. If
the factor ordering is not distorted by the presence of a varying speed of
light, one can further simplify the above equation to obtain%
\begin{equation}
\left(  -\frac{\partial^{2}}{\partial a^{2}}-\frac{q}{a}\frac{\partial
}{\partial a}+U_{c}\left(  a\right)  \right)  \Psi\left(  a\right)
=0,\label{WDWg}%
\end{equation}
where we have set $N=1$ and the quantum potential is defined as%
\begin{equation}
U_{c}\left(  a\right)  =\left(  \frac{3\pi}{2G\hbar}\right)  ^{2}a^{2}%
c^{6}\left(  t\right)  \left(  1-\frac{\Lambda}{3}a^{2}\right)  =\left(
\frac{3\pi c_{0}^{3}}{2G\hbar a_{0}^{3\alpha}}\right)  ^{2}a^{2+6\alpha
}\left(  1-\frac{\Lambda}{3}a^{2}\right)  .\label{Uac}%
\end{equation}
Note that the potential $U_{c}\left(  a\right)  $ vanishes in the same points
where $U\left(  a\right)  $ has its roots. Now, we are ready to discuss the
analogue of Eq.$\left(  \ref{WDW_1}\right)  $ in presence of a VSL
distortion\footnote{Note that for the special case $\alpha=-2/3$, one finds%
\begin{equation}
\left(  -\frac{\partial^{2}}{\partial a^{2}}-\frac{q}{a}\frac{\partial
}{\partial a}+\left(  \frac{3\pi c_{0}^{3}}{2G\hbar a_{0}^{-2}}\right)
^{2}\left(  a^{-2}-\frac{\Lambda}{3}\right)  \right)  \Psi\left(  a\right)
=0,
\end{equation}
that it means%
\begin{equation}
\left(  -\frac{\partial^{2}}{\partial a^{2}}-\frac{q}{a}\frac{\partial
}{\partial a}+\frac{K^{2}}{a^{2}}\right)  \Psi\left(  a\right)  =\frac{\Lambda
K^{2}}{3}\Psi\left(  a\right)  ,
\end{equation}
where we have defined $K=3\pi a_{0}^{2}/\left(  2l_{P}^{2}\right)  $. This
equation has exact solution in the form of a superposition of Bessel functions
$J_{\nu}\left(  x\right)  $ and $Y_{\nu}\left(  x\right)  $. However to obtain
eigenvalues one has to impose a large but finite boundary where the Bessel
functions vanish.}. To this purpose Eq.$\left(  \ref{WDWg}\right)  $ can be
cast into the form%
\begin{equation}
\frac{\int\mathcal{D}aa^{q}\Psi^{\ast}\left(  a\right)  \left[  -\frac
{\partial^{2}}{\partial a^{2}}-\frac{q}{a}\frac{\partial}{\partial a}+\left(
\frac{3\pi}{2l_{P}^{2}a_{0}^{3\alpha}}\right)  ^{2}a^{2+6\alpha}\right]
\Psi\left(  a\right)  }{\int\mathcal{D}aa^{q+4+6\alpha}\Psi^{\ast}\left(
a\right)  \Psi\left(  a\right)  }=3\Lambda\left(  \frac{\pi}{2l_{P}^{2}%
a_{0}^{3\alpha}}\right)  ^{2},\label{VeV}%
\end{equation}
where we have defined\textbf{ }$l_{P}=\sqrt{G\hbar/c_{0}^{3}}$. Because of the
VSL distortion, the asymptotic behavior of the trial wave function must be
different compared to $\left(  \ref{WDW_1}\right)  $. Since%
\begin{equation}
\left\{
\begin{array}
[c]{cc}%
K_{\nu}{\left(  x\right)  \rightarrow}\sqrt{\pi/\left(  2x\right)
}\mathrm{\exp}\left(  -x\right)   & x\rightarrow\infty\\
& \\%
\begin{array}
[c]{c}%
K_{0}{\left(  x\right)  \rightarrow-\ln}\left(  x\right)  \\
K_{\nu}{\left(  x\right)  \rightarrow}\Gamma\left(  \nu\right)  \left(
x/2\right)  ^{-\nu}/2
\end{array}
& x\rightarrow0
\end{array}
\right.  ,
\end{equation}
we find extremely useful the following assumption of the trial wave function%
\begin{equation}
\Psi\left(  a\right)  ={a}^{-\frac{q+1}{2}}\left(  \beta{a}\right)
^{-3\alpha}\exp\left(  -\frac{\beta{a}^{4}}{2}\right)  ,\label{Psib}%
\end{equation}
which is a small variation of $\left(  \ref{Psi}\right)  $. The exponential
encodes the large $a$ behavior, while $\left(  \beta{a}\right)  ^{-3\alpha}$
encodes the small scale factor behavior and $\beta$ is the variational
parameter. Plugging $\left(  \ref{Psib}\right)  $ into Eq.$\left(
\ref{VeV}\right)  $, after an integration over the scale factor $a\left(
t\right)  $, we find%
\begin{equation}
\Lambda_{q,\alpha}\left(  \beta\right)  =3\left(  \frac{2l_{P}^{2}%
a_{0}^{3\alpha}}{3\pi}\right)  ^{2}\left(  C_{q}\left(  \alpha\right)
{{\beta}^{\frac{3\left(  1+\alpha\right)  }{2}}}+\left(  \frac{3\pi}%
{2l_{P}^{2}a_{0}^{3\alpha}}\right)  ^{2}\sqrt{{{\beta\pi}}}\right)
,\label{LB}%
\end{equation}
where%
\begin{equation}
C_{q}\left(  \alpha\right)  =\frac{1}{4}{({q}^{2}-24\alpha-2\,q-7)\Gamma
\left(  {-\frac{1+3{\alpha}}{2}}\right)  }.
\end{equation}
We demand that%
\begin{equation}
\frac{d\Lambda_{q,\alpha}\left(  \beta\right)  }{d\beta}=\frac{3}{2{{\beta
}^{\frac{1}{{2}}}}}\left(  \frac{2l_{P}^{2}a_{0}^{3\alpha}}{3\pi}\right)
^{2}\left(  3\left(  1+\alpha\right)  C_{q}\left(  \alpha\right)  {{\beta
}^{\frac{2+3\alpha}{2}}}+\left(  \frac{3\pi}{2l_{P}^{2}a_{0}^{3\alpha}%
}\right)  ^{2}\sqrt{\pi}\right)  =0,\label{dLB}%
\end{equation}
where%
\begin{equation}
\bar{\beta}_{q}\left(  \alpha\right)  =\left(  -\left(  \frac{3\pi}{2l_{P}%
^{2}a_{0}^{3\alpha}}\right)  ^{2}\frac{\sqrt{\pi}}{3\left(  1+\alpha\right)
C_{q}\left(  \alpha\right)  }\right)  ^{\frac{2}{2+3\alpha}}.\label{betaSol}%
\end{equation}
with the conditions $1+\alpha\neq0$ and $2+3\alpha\neq0$. Plugging $\bar
{\beta}$ into $\Lambda_{q,\alpha}\left(  \beta\right)  $, one finds%
\begin{equation}
\Lambda_{q,\alpha}\left(  \beta\right)  =\bar{\beta}_{q}^{\frac{1}{2}}\left(
\alpha\right)  \sqrt{{{\pi}}}\frac{2+3{\alpha}}{{1+\alpha}}.\label{LB0}%
\end{equation}
The result is again dependent by the VSL\ parameter and on the reference scale
$a_{0}$. To this purpose we assume, without a loss of generality, that
$a_{0}=kl_{P}$. Then one gets%
\begin{equation}
\Lambda_{q}\left(  \alpha\right)  =l_{P}^{-2}\left(  -\left(  \frac{3\pi
}{2k^{3\alpha}}\right)  ^{2}\frac{\sqrt{\pi}}{3\left(  1+\alpha\right)
C_{q}\left(  \alpha\right)  }\right)  ^{\frac{1}{2+3\alpha}}\sqrt{{{\pi}}%
}\frac{2+3{\alpha}}{{1+\alpha}}.\label{Lq}%
\end{equation}
To have one and only one solution, we find a stationary point for $\Lambda
_{q}\left(  \alpha\right)  $ and we impose that%
\begin{equation}
\frac{d\Lambda_{q}\left(  \alpha\right)  }{d\alpha}=0\Longleftrightarrow
k=A\left(  q,\alpha\right)  \exp\left(  B\left(  q,\alpha\right)  \right)
\label{LambdaA}%
\end{equation}
where%
\begin{equation}
A\left(  q,\alpha\right)  =\frac{\pi^{\frac{5}{8}}}{3^{\frac{1}{4}}}\left(
{\left(  1+\alpha\right)  \Gamma\left(  -\frac{1+3\alpha}{2}\right)  \left(
-{q}^{2}+24\alpha-2q-7\right)  }\right)  ^{\frac{1}{4}}%
\end{equation}
and%
\begin{equation}
B\left(  q,\alpha\right)  =\frac{{(3\alpha{q}^{2}-72{\alpha}^{2}+6\alpha
q+2{q}^{2}-27\alpha+4q+14)\,\Psi\left(  -\frac{1+3\alpha}{2}\right)
+48\alpha+32}}{8({q}^{2}-24\alpha+2q+7)}.
\end{equation}
In the table below, it is shown the result of the procedure $\left(
\ref{LambdaA}\right)  $ for some specific choices of $q$%
\begin{equation}
\left\{
\begin{tabular}
[c]{|c|c|c|}\hline
$q=1$ & $k_{0}=0.5779378002$ & $\bar{\alpha}=-2.007150679$\\\hline
$q=0$ & $k_{0}=0.5843673484$ & $\bar{\alpha}=-1.988596177$\\\hline
$q=-1$ & $k_{0}=0.6030705325$ & $\bar{\alpha}=-1.940190188$\\\hline
\end{tabular}
\ \ \ \ \right.  .
\end{equation}
\begin{figure}[h]
\centering\includegraphics[width=2.8in]{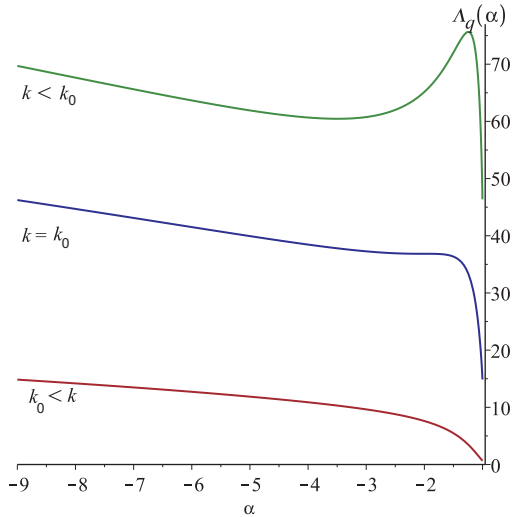}\caption{Plot of
$\Lambda_{q}\left(  \alpha\right)  $ as a function of $\alpha$ depicted for
$q=1$. The local minimum and the local maximum appear below the critical scale
$k_{0}$. For $k=k_{0}$ there is only a stationay point which disappears for
$k>k_{0}$.}%
\label{Lambda}%
\end{figure}As depicted in Fig.$\left(  \ref{Lambda}\right)  $, the couple
$\left(  k_{0},\bar{\alpha}\right)  $ does not represent the solution of the
problem, because the point is stationary and not a local minimum. Rather we
can interpret the couple $\left(  k_{0},\bar{\alpha}\right)  $ as a critical
value below which a minimum and a maximum appear. In particular, as shown in
Fig..$\left(  \ref{Lambda}\right)  $, for $k<k_{0}$ and $\alpha<\bar{\alpha}$
we have a minimum and for $k<k_{0}$ and $\alpha>\bar{\alpha}$, we have a
maximum. In the spirit of the variational procedure only the minimum can be
considered as the solution of the problem. Note that the lower the value of
$k_{0}$, the higher the value of $\Lambda_{q}\left(  \bar{\alpha}\right)  $.
Note also that the value of $k_{0}$ is transplanckian. From the expression of
$\Lambda_{q}\left(  \alpha\right)  $, this is true when ${\alpha>0}$ or when
${\alpha<-2/3}$, otherwise when ${-2/3<\alpha<0}$, the behavior on $k$
reverses. From the assumption $\left(  \ref{c(t)}\right)  $, we see that for
${\alpha>0}$, we have $c\left(  t\right)  \gg c_{0}$ when the scale factor
$a\left(  t\right)  \gg a_{0}$ and this is ruled out by observation.
Therefore, the correct range of solutions is when ${\alpha<-2/3}$. Furthermore
to have also positive solutions we need ${\alpha<-1}$. The physical reason of
why we obtain solutions in the negative range of ${\alpha}$ is that in the
early universe one expects to measure strong quantum effects when the scale
factor $a\left(  t\right)  $ is really small. In this framework, this is
realized with a speed of light which is really big compared to $c_{0}$. It is
interesting to note that for $k>k_{0}$, there is no solution at all and for
$k\gg k_{0}$, $\Lambda_{q}\left(  \alpha\right)  \ll1$ in Planck's units and
$\alpha\in\left(  \bar{\alpha}-\varepsilon,\bar{\alpha}\right)  $ with
$\varepsilon>0$ arbitrarily small.

\section{Conclusions}

\label{p3}Motivated by the results obtained in different contexts of distorted
gravity and in special way in the HL theory\cite{RemoHL}, in this paper we
have examined the possibility that the cosmological constant can be considered
as an eigenvalue of an appropriate Sturm-Liouville problem of a VSL theory.
This interpretation is not new and it has been explored in different
contexts\cite{GMLM,GMLM1,RemoHL}. What is different in this paper is that the
WDW equation on a FLRW background in a mini-superspace approach reveals a
complete analogy with a Sturm-Liouville problem and the cosmological constant
has the natural interpretation of its related eigenvalue. The WDW equation has
been examined taking account of the factor order ambiguity. We have probed
ordinary gravity without matter fields with different families of trial wave
functions and we have found no sign of a cosmological constant induced by
quantum fluctuations. Of course we have not exhausted all the possible choices
of the trial wave functions. However, the form we have adopted has been chosen
using the standard criteria for a variational approach. Therefore, we can
conjecture that for a mini-superspace approach without matter fields, a
cosmological constant cannot be generated. The introduction of a VSL
\begin{equation}
c\left(  t\right)  =c_{0}\left(  \frac{a\left(  t\right)  }{a_{0}}\right)
^{\alpha}%
\end{equation}
makes the situation different because of the power law on the scale factor.
This modification is also supported by the following alternative definition of
the speed of light%
\begin{equation}
c\left(  E/E_{\mathrm{Pl}}\right)  =\frac{dE}{dp}=c_{0}\frac{g_{2}\left(
E/E_{\mathrm{Pl}}\right)  }{g_{1}\left(  E/E_{\mathrm{Pl}}\right)
},\label{c(E)}%
\end{equation}
which can be easily extracted if one introduces Gravity's Rainbow into the
FLRW metric. In this formulation, the space-time geometry is described by the
deformed metric
\begin{equation}
ds^{2}=-\frac{N^{2}\left(  t\right)  }{g_{1}^{2}\left(  E/E_{\mathrm{Pl}%
}\right)  }dt^{2}+\frac{a^{2}\left(  t\right)  }{g_{2}^{2}\left(
E/E_{\mathrm{Pl}}\right)  }d\Omega_{3}^{2}~,\label{FLRWMod}%
\end{equation}
where $g_{1}(E/E_{\mathrm{Pl}})$ and $g_{2}(E/E_{\mathrm{Pl}})$ are functions
of energy, which incorporate the deformation of the metric. Concerning the
low-energy limit it is required to consider
\begin{equation}
\lim_{E/E_{\mathrm{Pl}}\rightarrow0}g_{1}\left(  E/E_{\mathrm{Pl}}\right)
=1\qquad\mathrm{and}\qquad\lim_{E/E_{\mathrm{Pl}}\rightarrow0}g_{2}\left(
E/E_{\mathrm{Pl}}\right)  =1,
\end{equation}
and thus to recover the usual background $\left(  \ref{FRW}\right)  $. Hence,
$E$ quantifies the energy scale at which quantum gravity effects become
apparent. For instance, one of these effects would be that the graviton
distorts the background metric as we approach the Planck scale. In a distorted
FLRW metric the dispersion relation for a massless graviton is%
\begin{equation}
E^{2}g_{1}^{2}\left(  \frac{E}{E_{P}}\right)  =p^{2}g_{2}^{2}\left(  \frac
{E}{E_{P}}\right)  ,
\end{equation}
leading to $\left(  \ref{c(E)}\right)  $. Setting for example%
\begin{align}
g_{1}\left(  E/E_{\mathrm{Pl}}\right)   &  =1\nonumber\\
g_{2}\left(  E/E_{\mathrm{Pl}}\right)   &  =1+\left(  \frac{a\left(  t\right)
}{a_{0}}\right)  ^{\alpha},\label{VSLb}%
\end{align}
one obtains a different, but equivalent form of the VSL. This formulation has
the advantage to avoid technical complications as in Ref.\cite{RGES}. The
choice in $\left(  \ref{VSLb}\right)  $ appears to be connected also to the
following potential%
\begin{equation}
a^{4}\left(  t\right)  \left[  \frac{6}{a^{2}\left(  t\right)  }-\frac{96\pi
Gb}{a^{4}\left(  t\right)  }-\frac{\allowbreak3456\pi^{2}G^{2}c}{a^{6}\left(
t\right)  }-2\Lambda\right]  ,\label{HL}%
\end{equation}
coming from a HL theory without detailed balanced condition. In this kind of
potential, one discovers positive eigenvalues depending on the various
coupling constants choices. However the potential $\left(  \ref{HL}\right)  $
appears to be more flexible to produce positive eigenvalues. It is for this
reason that the structure of the trial wave function we have used in this
paper is more elaborated compared to a simple gaussian function which has bees
used in a HL theory\cite{RemoHL}. The procedure of finding a minimum for
$\Lambda\left(  \beta\right)  $ of Eq.$\left(  \ref{LB}\right)  $ has produced
a result depending on two other parameters, the power $\alpha$ and the
reference scale $k$. A further minimization procedure allows to select one
value compatible with the procedure which however does not constitute the
final answer, rather it has been interpreted as a critical value below which
we have eigenvalues while above which we have none. Note that the appearance
of eigenvalues compatible with the procedure is in the transplanckian regime
and for negative values of $\alpha$. Negative values of $\alpha$ have been
found in Ref.\cite{harko}, even if the authors discuss the \textquotedblleft
Creation from Nothing\textquotedblright\ problem. Note that for Planckian and
cisplanckian values of the scale $a_{0}$, the eigenvalue does not appear and
for larger scales, like the inflationary one, the whole expression in $\left(
\ref{Lq}\right)  $ becomes very small for every value of $\alpha<-1$. At this
stage of calculation, we do not know if this behavior is simply a failure of
the approach or further information can be extracted.

\appendix{}

\section{Integrals for $\Psi\left(  a\right)  =\exp\left(  -\frac{\beta{a}%
^{2}}{2}\right)  U{\left(  \frac{q+1}{4},\frac{q+1}{2},\beta{a}^{2}\right)  }%
$}

\label{Appe1}If the trial wave function assumes the form%
\begin{equation}
\Psi\left(  a\right)  =\exp\left(  -\frac{\beta{a}^{2}}{2}\right)  U{\left(
\frac{q+1}{4},\frac{q+1}{2},\beta{a}^{2}\right)  ,}%
\end{equation}
then, for practical purposes, it can be transformed into%
\begin{equation}
\Psi\left(  a\right)  ={\frac{\left(  \beta{a}^{2}\right)  ^{\left(
1-q\right)  /4}}{\sqrt{\pi}}K_{\left(  q-1\right)  /4}{\left(  \frac{\beta
{a}^{2}}{2}\right)  },} \label{BesselK}%
\end{equation}
where we have used the identity $\left(  \ref{Kummer}\right)  $%
\begin{equation}
U{\left(  a+1/2,2a+1,2x\right)  =}\mathrm{\exp}\left(  x\right)  K_{a}{\left(
x\right)  /}\left(  \sqrt{\pi}\left(  2x\right)  ^{a}\right)  ,
\end{equation}
with $a=\left(  q-1\right)  /4$ and $x=\beta{a}^{2}/2$. Plugging the trial
wave function $\left(  \ref{BesselK}\right)  $ into the kinetic term, one gets%
\begin{align}
&  \left[  -\frac{\partial^{2}}{\partial a^{2}}-\frac{q}{a}\frac{\partial
}{\partial a}\right]  \Psi\left(  a\right) \nonumber\\
&  =-\beta^{2}a^{2}\exp\left(  -\frac{\beta{a}^{2}}{2}\right)  U{\left(
\frac{q+1}{4},\frac{q+1}{2},\beta{a}^{2}\right)  .} \label{HPsi}%
\end{align}
If we multiply the expression $\left(  \ref{HPsi}\right)  $ on the left by
$a^{q}\Psi^{\ast}\left(  a\right)  $ and we integrate over the scale factor,
we find%
\begin{gather}
\int_{0}^{\infty}daa^{q}\Psi^{\ast}\left(  a\right)  \left[  -\frac
{\partial^{2}}{\partial a^{2}}-\frac{q}{a}\frac{\partial}{\partial a}\right]
\Psi\left(  a\right)  =-\beta^{2}\int_{0}^{\infty}daa^{q+2}\exp\left(
-\beta{a}^{2}\right)  U^{2}{\left(  \frac{q+1}{4},\frac{q+1}{2},\beta{a}%
^{2}\right)  }\nonumber\\
=-\,\frac{\beta^{\frac{5-q}{2}}}{\pi}\int_{0}^{\infty}daa^{3}K_{\left(
q-1\right)  /4}^{2}{\left(  \frac{\beta{a}^{2}}{2}\right)  }\nonumber\\
{\underset{\sqrt{\beta/2}a=\sqrt{x}}{=}-}\frac{{2}}{\pi}\beta^{\frac{1-q}{2}%
}\int_{0}^{\infty}dxxK_{\left(  q-1\right)  /4}^{2}{\left(  x\right)  }%
=-\frac{\beta^{\frac{1-q}{2}}}{\pi}\Gamma\left(  \frac{3+q}{4}\right)
\Gamma\left(  \frac{5-q}{4}\right)  , \label{HP1}%
\end{gather}
where we have used the following relationship\cite{PBM}%
\begin{gather}
\int_{0}^{\infty}dxx^{\alpha-1}K_{\mu}{\left(  cx\right)  }K_{\nu}{\left(
cx\right)  }\nonumber\\
=\frac{2^{\alpha-3}}{c^{\alpha}\Gamma\left(  \alpha\right)  }\Gamma\left(
\frac{\alpha+\mu+\nu}{2}\right)  \Gamma\left(  \frac{\alpha+\mu-\nu}%
{2}\right)  \Gamma\left(  \frac{\alpha+\nu-\mu}{2}\right)  \Gamma\left(
\frac{\alpha-\mu-\nu}{2}\right) \nonumber\\
\operatorname{Re}c>0;\operatorname{Re}\alpha>\left\vert \operatorname{Re}%
\nu\right\vert +\left\vert \operatorname{Re}\mu\right\vert .
\end{gather}
and where $\Gamma\left(  x\right)  $ is the gamma function. The contribution
coming from the potential term without the VSL distortion is composed by{%
\begin{gather}
\int_{0}^{\infty}daa^{q}\Psi^{\ast}\left(  a\right)  a^{2}\Psi\left(
a\right)  =\int_{0}^{\infty}daa^{q+2}\exp\left(  -\beta{a}^{2}\right)
U^{2}{\left(  \frac{q+1}{4},\frac{q+1}{2},\beta{a}^{2}\right)  }\nonumber\\
=\frac{\beta^{\frac{1-q}{2}}}{\pi}\int_{0}^{\infty}daa^{3}K_{\left(
q-1\right)  /4}^{2}{\left(  \frac{\beta{a}^{2}}{2}\right)  }\nonumber\\
{\underset{\sqrt{\beta/2}a=\sqrt{x}}{=}2}\frac{\beta^{-\frac{3+q}{2}}}{\pi
}\int_{0}^{\infty}dxxK_{\left(  q-1\right)  /4}^{2}{\left(  x\right)  }%
=\frac{1}{\pi\beta^{\frac{3+q}{2}}}\Gamma\left(  \frac{3+q}{4}\right)
\Gamma\left(  \frac{5-q}{4}\right)  \label{d}%
\end{gather}
and%
\begin{gather}
\int_{0}^{\infty}daa^{q}\Psi^{\ast}\left(  a\right)  a^{4}\Psi\left(
a\right)  =\int_{0}^{\infty}daa^{q+4}\exp\left(  -\beta{a}^{2}\right)
U^{2}{\left(  \frac{q+1}{4},\frac{q+1}{2},\beta{a}^{2}\right)  }\nonumber\\
=\frac{\beta^{\frac{1-q}{2}}}{\pi}\int_{0}^{\infty}daa^{5}K_{\left(
q-1\right)  /4}^{2}{\left(  \frac{\beta{a}^{2}}{2}\right)  }\nonumber\\
{\underset{\sqrt{\beta/2}a=\sqrt{x}}{=}4}\frac{\beta^{-\frac{5+q}{2}}}{\pi
}\int_{0}^{\infty}dxx^{2}K_{\left(  q-1\right)  /4}^{2}{\left(  x\right)
=\frac{1}{2\beta^{\frac{5+q}{2}}}\Gamma\left(  \frac{5+q}{4}\right)
\Gamma\left(  \frac{7-q}{4}\right)  }. \label{e}%
\end{gather}
The contribution coming from the potential term with the VSL distortion is
composed by%
\begin{gather}
\int_{0}^{\infty}daa^{q}\Psi^{\ast}\left(  a\right)  a^{2+6\alpha}\Psi\left(
a\right)  =\int_{0}^{\infty}daa^{q+2+6\alpha}\exp\left(  -\beta{a}^{2}\right)
U^{2}{\left(  \frac{q+1}{4},\frac{q+1}{2},\beta{a}^{2}\right)  }\nonumber\\
=\frac{\beta^{\frac{1-q}{2}}}{\pi}\int_{0}^{\infty}daa^{3+6\alpha}K_{\left(
q-1\right)  /4}^{2}{\left(  \frac{\beta{a}^{2}}{2}\right)  \underset
{\sqrt{\beta/2}a=\sqrt{x}}{=}}\frac{{2}^{1+3\alpha}}{\pi\beta^{2+3\alpha
+\frac{q-1}{2}}}\int_{0}^{\infty}dxx^{1+3\alpha}K_{\left(  q-1\right)  /4}%
^{2}{\left(  x\right)  }\nonumber\\
=\frac{{2}^{6\alpha}}{\pi\beta^{\frac{3+q}{2}+3\alpha}\Gamma\left(
2+3\alpha\right)  }\Gamma\left(  \frac{3+6\alpha+q}{4}\right)  \Gamma
^{2}\left(  \frac{2+3\alpha}{2}\right)  \Gamma\left(  \frac{5+6\alpha-q}%
{4}\right)  \label{f}%
\end{gather}
and%
\begin{gather}
\int_{0}^{\infty}daa^{q}\Psi^{\ast}\left(  a\right)  a^{4+6\alpha}\Psi\left(
a\right)  =\int_{0}^{\infty}daa^{q+4+6\alpha}\exp\left(  -\beta{a}^{2}\right)
U^{2}{\left(  \frac{q+1}{4},\frac{q+1}{2},\beta{a}^{2}\right)  }\nonumber\\
=\frac{\beta^{\frac{1-q}{2}}}{\pi}\int_{0}^{\infty}daa^{5+6\alpha}K_{\left(
q-1\right)  /4}^{2}{\left(  \frac{\beta{a}^{2}}{2}\right)  \underset
{\sqrt{\beta/2}a=\sqrt{x}}{=}}\frac{2^{2+3\alpha}}{\pi\beta^{3+3\alpha
+\frac{q-1}{2}}}\int_{0}^{\infty}dxx^{2+3\alpha}K_{\left(  q-1\right)  /4}%
^{2}{\left(  x\right)  }\nonumber\\
{=\frac{2^{2+6\alpha}}{\pi\beta^{\frac{5+q}{2}+3\alpha}\Gamma\left(
3+3\alpha\right)  }\Gamma\left(  \frac{5+6\alpha+q}{4}\right)  \Gamma}%
^{2}{\left(  \frac{3+3\alpha}{2}\right)  \Gamma\left(  \frac{7+6\alpha-q}%
{4}\right)  }. \label{g}%
\end{gather}
}

\section{Integrals for $\Psi\left(  a\right)  ={a}^{-\frac{q+1}{2}}\left(
\beta{a}\right)  ^{\frac{\rho}{2}}\exp\left(  -\frac{\beta{a}^{\lambda}}%
{2}\right)  $}

\label{Appe2}If the trial wave function assumes the form%
\begin{equation}
\Psi\left(  a\right)  ={a}^{-\frac{q+1}{2}}\left(  \beta{a}\right)
^{\frac{\rho}{2}}\exp\left(  -\frac{\beta{a}^{\lambda}}{2}\right)  {,}%
\end{equation}
when we plug the trial wave function $\left(  \ref{BesselK}\right)  $ into the
kinetic term, one gets%
\begin{align}
&  \left[  -\frac{\partial^{2}}{\partial a^{2}}-\frac{q}{a}\frac{\partial
}{\partial a}\right]  \Psi\left(  a\right)  =\frac{1}{4}\,\exp\left(
-\frac{\beta{a}^{\lambda}}{2}\right)  \beta^{\frac{\rho}{2}}{a}^{\frac
{\rho-5-q}{2}}\left[  \left(  {q}^{2}-{\rho}^{2}+4\rho-2q-3\right)  \right.
\nonumber\\
&  =\left.  2{\beta a}^{\lambda}\left(  {\lambda}^{2}+\rho\,\lambda
-2\lambda\right)  -{a}^{2\,\lambda}{\beta}^{2}{\lambda}^{2}\right]  {.}%
\end{align}
If we multiply the expression $\left(  \ref{HPsi}\right)  $ on the left by
$a^{q}\Psi^{\ast}\left(  a\right)  $ and we integrate over the scale factor,
we find%
\begin{equation}
\int_{0}^{\infty}daa^{q}\Psi^{\ast}\left(  a\right)  \left[  -\frac
{\partial^{2}}{\partial a^{2}}-\frac{q}{a}\frac{\partial}{\partial a}\right]
\Psi\left(  a\right)  ={\frac{\lambda\rho+{q}^{2}-2\lambda-2\,q+1}{4\lambda
}\Gamma\left(  {\frac{-2+\rho}{\lambda}}\right)  {\beta}^{{\frac{\lambda
\rho-\rho+2}{\lambda}}}},
\end{equation}
where $\Gamma\left(  x\right)  $ is the gamma function. {The contribution
coming from the potential term with the VSL distortion is composed by%
\begin{gather}
\int_{0}^{\infty}daa^{q}\Psi^{\ast}\left(  a\right)  a^{2+6\alpha}\Psi\left(
a\right)  =\int_{0}^{\infty}daa^{1+6\alpha+\rho}\beta^{\rho}\exp\left(
-\beta{a}^{\lambda}\right) \nonumber\\
={\frac{1}{\lambda}{\beta}^{{\frac{\lambda\rho-6\alpha-\rho-2}{\lambda}}%
}\Gamma\left(  {\frac{2+6\alpha+\rho}{\lambda}}\right)  }%
\end{gather}
and%
\begin{gather}
\int_{0}^{\infty}daa^{q}\Psi^{\ast}\left(  a\right)  a^{4+6\alpha}\Psi\left(
a\right)  =\int_{0}^{\infty}daa^{3+6\alpha+\rho}\beta^{\rho}\exp\left(
-\beta{a}^{\lambda}\right) \nonumber\\
={\frac{1}{\lambda}{\beta}^{{\frac{\lambda\rho-6\alpha-\rho-4}{\lambda}}%
}\Gamma\left(  {\frac{4+6\alpha+\rho}{\lambda}}\right)  }%
\end{gather}
}


\begin{thebibliography}{99}                                                                                               %


\bibitem {DeWitt}B. S. DeWitt, \textsl{Phys. Rev.} \textbf{160}, 1113 (1967).

\bibitem {ADM}R. Arnowitt, S. Deser, and C. W. Misner, in \textit{Gravitation:
An Introduction to Current Research,} edited by L. Witten (John Wiley \& Sons,
Inc., New York, 1962); B. S. DeWitt, Phys. Rev. \textbf{160}, 1113 (1967). arXiv:gr-qc/0405109.

\bibitem {f(R)}S. Capozziello and M.De Laurentis, \textsl{Phys. Rept.}
\textbf{509}, 167 (2011); arXiv:1108.6266 [gr-qc]. S. Nojiri and S. D.
Odintsov, \textsl{Phys. Rept.} \textbf{505}, 59 (2011); arXiv:1011.0544
[gr-qc]. T. P. Sotiriou and V. Faraoni, \textsl{Rev. Mod. Phys.} \textbf{82},
451 (2010); arXiv:0805.1726 [gr-qc].

\bibitem {Horava}P. H\v{o}rava, \textsl{JHEP}, 0903, \textbf{020} (2009).
ArXiv: 0812.4287 [hep-th]; P. H\v{o}rava, \textsl{Phys. Rev. Lett.}
\textbf{102}, 161301 (2009) ArXiv: 0902.3657 [hep-th].

\bibitem {Lifshitz}E.M. Lifshitz, \textsl{Zh. Eksp. Toer. Fiz.} \textbf{11},
255; 269 (1941).

\bibitem {MagSmo}J. Magueijo and L. Smolin, \textsl{Class. Quant. Grav.}
\textbf{21}, 1725 (2004) [arXiv:gr-qc/0305055].

\bibitem {AM}S.\ Alexander and J.\ Magueijo, \textsl{Noncommutative geometry
as a realization of varying speed of light cosmology} in Proceedings of the
XIIIrd Rencontres de Blois, pp281, 2004 [arXiv:hep-th/0104093].

\bibitem {GMLM}R. Garattini and G. Mandanici, \textsl{Phys. Rev.} \textbf{D
83} 084021 (2011); [arXiv:1102.3803 [gr-qc]]. R. Garattini, \textsl{JCAP} 1306
(2013)\textbf{ 017}; arXiv:1210.7760 [gr-qc]. R. Garattini and B. Majumder,
\textsl{Nucl. Phys.} \textbf{B 884} 125, (2014) [arXiv:1311.1747 [gr-qc]].

\bibitem {GMLM1}R. Garattini, \textsl{Phys.Lett.} \textbf{B} 685 329 (2010);
[arXiv:0902.3927 [gr-qc]]. R. Garattini and F. S.N. Lobo, \textsl{Phys. Rev.}
\textbf{D 85} 024043 (2012); [arXiv:1111.5729 [gr-qc]]. R. Garattini and G.
Mandanici, \textsl{Phys. Rev.} \textbf{D 85} 023507 (2012); [arXiv:1109.6563
[gr-qc]]. R. Garattini and B. Majumder, \textsl{Nucl. Phys.} \textbf{B 883}
598, (2014); [arXiv:1305.3390 [gr-qc]. R. Garattini and F. S. N. Lobo,
\textsl{Eur. Phys. J.} \textbf{C 74} 2884, (2014); [arXiv:1303.5566 [gr-qc]].

\bibitem {RemoHL}R. Garattini, \textsl{Phys. Rev.} \textbf{D 86} 123507
(2012); [arXiv:0912.0136 [gr-qc]].

\bibitem {OBCZ}O. Bertolami and C. A. D. Zarro, \textsl{Phys. Rev.} \textbf{D
84} 044042, (2011); [arXiv:1106.0126 [hep-th]]

\bibitem {AAEEAY}A. V. Astashenok, E. Elizalde and A. V. Yurov,
\textsl{Astrophys. Space Sci.} 349 25 (2014);[arXiv:1212.4268 [astro-ph.CO]].

\bibitem {RGES}R. Garattini and E. N. Saridakis, \textit{Gravity's Rainbow: a
bridge towards Horava-Lifshitz gravity}, [arXiv:1411.7257 [gr-qc]].

\bibitem {harko}T. Harko and M. K. Mak, \textit{Class. Quantum Grav.}
\textbf{16}, 2741 (1999).

\bibitem {moffat}J. W. Moffat, \textsl{Int. J. Mod. Phys. D} \textbf{2} 351, (1993).

\bibitem {Albrecht}A. Albrecht and J. Magueijo, \textsl{Phys. Rev. D}
\textbf{59}, 043516 (1999).

\bibitem {Barrow1}J. D. Barrow, \textsl{Phys. Rev. D} \textbf{59},043515 (1999).

\bibitem {Barrow2}J. D. Barrow and J. Magueijo, \textsl{Phys. Lett. B}
\textbf{443}, 104 (1999).

\bibitem {Barrow3}J. D. Barrow and J. Magueijo, \textsl{Phys. Lett. B}
\textbf{447},246 (1999).

\bibitem {Kolb90}E. W. Kolb and M. S. Turner, The Early Universe (Wiley, New
York, 1990).

\bibitem {guth}A.H. Guth, \textit{Phys. Rev. D} \textbf{23}, 347 (1981).

\bibitem {linde}A. Linde, \textit{Phys. Lett. B} \textbf{108}, 389 (1982).

\bibitem {AA}A. Albrecht and P. Steinhardt, \textit{Phys. Rev. Lett.}
\textbf{48}, 1220 (1982).

\bibitem {veneziano}G. Veneziano, \textit{Phys. Lett. B} \textbf{406}, 297 (1997).

\bibitem {ellis}G.F.R. Ellis, \textit{Gen. Rel. Grav.}, \textbf{39}, 511 (2007).

\bibitem {magueio}J. Magueijo, \textit{Phys. Rev. D} \textbf{62}, 103521 (2000).

\bibitem {moff}J. Magueijo, J. W. Moffat, \textit{Gen.Rel.Grav.} \textbf{40},
1797 (2008).

\bibitem {Vilenkin}A.\ Vilenkin, Phys.\ Rev. \textbf{D 37} (1988) 888.

\bibitem {HH}J.\ B.\ Hartle and S.\ W.\ Hawking, Phys.\ Rev. \textbf{D 28}
(1983) 2960; S.\ W.\ Hawking, Nucl.\ Phys. \textbf{B 239} (1984) 257.

\bibitem {harko1}T. Harko, H. Q. Lu, M. k. Mak and K. S. Cheng,
\textit{Europhys. Lett.} \textbf{49}, 814 (2000).

\bibitem {sho}F. Shojai, S. Molladavoudi, \textit{Gen. Relativ. Gravit.}
\textbf{39}, 795 (2007).



\bibitem {VSL}. J. W. Moffat, Published in \textquotedblleft\textit{Janke, W.
(ed.) et al.: Fluctuating paths and fields}\textquotedblright\ 741-757;
preprint astro-ph/9811390 (1998). ;

\bibitem {VSLQC}T. Harko, H. Q. Lu, M. K. Mak and K. S. Cheng,
\textsl{Europhys. Lett.}, \textbf{49} (\textbf{6}), pp. 814--820 (2000). A.V.
Yurov and V.A. Yurov, \textquotedblleft\textit{The semiclassical tunneling
probability in quantum cosmologies with varying speed of light}%
\textquotedblright, arXiv:hep-th/0410231.

\bibitem {MagSmo1}J. Magueijo and L. Smolin, \textsl{Phys. Rev.} \textbf{D 67}
(2003) 044017 [arXiv:gr-qc/0207085].

\bibitem {Wiltshire}D. L. Wiltshire, An introduction to quantum cosmology.
[gr-qc/0101003]. D. L. Wiltshire, \textsl{Gen.Rel.Grav.} \textbf{32}, 515
(2000); [arXiv: gr-qc/9905090]. N. Kontoleon and D.L. Wiltshire, \textsl{Phys.
Rev.} \textbf{D 59}, 063513 (1999); [arXiv: gr-qc/9807075].

\bibitem {WPKA}S. Watson, M. J. Perry, G. L. Kane and F. C. Adams,
\textsl{JCAP} 0711 017 (2007); [arXiv: hep-th/0610054].

\bibitem {PBM}A. P. Prudnikov, Yu. A. Brychkov and O.I. Marichev,
\textit{Integrals and Series,Vol. 2: More Special Functions}, edited by Gordon
and Breach Science Publishers, Second Printing 1998.


\end{thebibliography}
\end{document}